\documentclass[10pt,letterpaper]{article}
\usepackage{opex3}
\usepackage{color}
\usepackage{threeparttable}

\begin{document}

\title{Wavelet analysis and HHG in nanorings: their applications in logic gates and memory mass devices}
\author{Dario Cricchio and Emilio Fiordilino}
\email{dario.cricchio@unipa.it}
 \address{Dipartimento di Fisica e Chimica, Universit\`a di Palermo, Via Archirafi 36, 90123 Palermo (Italy)}

\begin{abstract}
We study the application of one nanoring driven by a  laser field in different states of polarization in logic circuits. In particular we show that assigning boolean values to different state of the incident laser field  and to the emitted signals, we can create logic gates such as OR, XOR and AND. We also show the possibility to make logic circuits such as half-adder and full-adder using one and two nanoring respectively. Using two nanorings we made the Toffoli gate. Finally we use the final angular momentum acquired by the electron to store information and hence show the possibility to use an array of  nanorings as a mass memory device.
\end{abstract}

\section{Introduction}
The rapid development of the nanosciences  opens new technological frontiers. Computer science is one of them. In fact by some years we are observing a saturation in the microprocessors performance which requires the finding of new objects that allow the construction of logic circuits faster and  smaller than today's. For these reasons the research is focusing on a new kind of devices that show electrical and optical properties and that can be suitable to construct logic circuits: nanorings and graphenes are an example of them. They show interesting behavior  in nanotechnology, optic and computer science  thanks to their  electrical and optical properties that make them apt to implementation in logic circuits. One important property of the graphene is its dispersion relation. In fact it is linear around the six corners of the hexagonal pattern leading to the vanishing of the effective mass of electrons and holes \cite{Wallace}.
Another  noteworthy object is the nanoring, a planar structure with a ring geometry.  Gold nanoring arrays are studied for biosensing applications in the fiber-optic communication window \cite{goldring}. 
The interaction between a strong laser field and molecules give the possibility to observe non linear processes that cannot be described by the standard perturbation theory. 

Nanorings present several optical properties, in particular they show that the  HHG spectrum, obtained when they are driven by a circularly polarized laser field, is more intense than the HHG spectrum obtained from benzene \cite{Baldea}.
Analytic and experimental studies on mesoscopic molecules show that the large size and polarizability make them interesting sources of harmonics \cite{Voronov, Alon, Averbukh, Petra, Songard}. In fact nanorings driven by a two color laser field can efficiently emit a broad harmonic spectrum; and in \cite{AutoLPL, AutoPRA2012, AutoPRA2013} it is shown the possibility of controlling the polarization
of the emitted harmonics by changing the laser parameters  and also to have a final angular momentum of the electron in the nanoring.
It is also numerically demonstrate that  ring current initial states generate a nonstandard ionization and harmonic generation by strong laser pulses
 and that the internal state is transferred to the tunneling wave packets, which receive an initial momentum \cite{xie}.
 
A circularly polarized ultraviolet laser pulse  excites a unidirectional valence-type electronic ring current in an oriented molecule. The net ring current generated by the laser pulse is about 84.5 $\mu$A and could  be induced by means of permanent magnetic fields \cite{barth}. magnetic fields can also induced by a circular polarized laser pulse \cite{barth2}.

The aromatic group, containing six carbon atoms, can be considered as the nanoring with the smallest radius. It corresponds to a single cell of graphene. Nanorings with large radius are obtained joining strips of graphene \cite{nanobook}, by self-assembling RNA nanorings based on RNAI/IIi kissing complexes \cite{rnaring} or by an array of nano apertures \cite{nanoaperture}.
A mesoscopic ring treated with a magnetic flux is used to make  NAND gate \cite{Meso}, where the ring is attached symmetrically to two  metallic electrodes and two gate voltages are applied in one arm of the ring, these are treated as the two inputs of the NAND gate.
Optical signals are used in  \cite{Resonator} where two silicon microring resonators are used to perform XOR and XNOR operations. In \cite{optcin} it is shown the possibility to generate  optical NOT and XOR logic operation based on terahertz optical asymmetric demultiplexer.
Also thermal logic gates exist  that can do simple calculations using the phonons as information \cite{logicp}. The symmetry of the wave packets and their shape are used like quantum bits \cite{wavebit}.  

This paper deals with the possibility to use nanorings, driven by a laser field, as  logical gates.
Our system is composed by one nanoring, with only one active electron, driven by an elliptically polarized laser field. We calculate the harmonics  and the Raman lines emitted and the angular momentum acquired by the electron in different states of polarization of the incident laser field. The process that we use in this work is the high harmonic generation (HHG) \cite{Ferray, dipiazzakaren}.
The HHG spectrum is generated by a non-linearity of the
electron-laser interaction energy that forces the system to emit a spectrum  of harmonics of the laser field \cite{Matos}. Several studies shows that the use of a two-color laser field or of a  laser photon energy resonant between the ground state and the first excited state can enhance the emitted spectrum \cite{AutoJOSB, ganeev1, 2color, Hinsche}.
Then we calculate the HHG spectra and, from the use of the wavelet transform, we recognize the presence  of a signal composed by the first two harmonics and the Raman lines. We perform our calculations by varying the states of polarization  of the incident laser field. We also calculate the angular momentum acquired by the electron in order to use it to create a truth table of logical operations and then to use the nanoring as a logic gate. We give examples of implementation  of nanoring to construct a basic logic circuit such as the half and the full-adder. We can also construct a reversible logic gate, the Toffoli gate, using two nanorings. Finally we discuss the possibility of constructing a memory mass device using an array of nanorings.
\section{Theory}
In our system we consider one nanoring of radius $R$ in the single active electron approximation driven by one laser elliptically polarized along the same plane. The equation of the electric field is:
\begin{eqnarray}\
\vec{\mathcal E}_L (t)&=&\mathcal E_0f(t) \bigl[\hat{\epsilon}_x \cos\left (\beta \right)\cos(\omega_Lt)+\nonumber\\
 &+&\hat{\epsilon}_y\sin\left (\beta \right)\sin(\omega_Lt)\bigr]
\end{eqnarray}
where $f(t)$ is the pulse shape,  $\hat{\epsilon}_x$ and $\hat{\epsilon}_y$ are the unit vectors along the $x$ and $y$ axes respectively, $\omega_L$ is the angular frequency of the external laser field and $\beta$ is a parameter that characterize the polarization of the electric field. In particular for $
\beta=0^\circ$ we have a laser polarized along the $x$ axis, for $
\beta=90^\circ$ we have a laser polarized along the $y$ axis and for $
\beta=45^\circ$ we have a circular polarized laser. 
The Hamiltonian of the system is
\begin{eqnarray}
\mathcal{H}&=& \frac{\hbar^2\ell_z^2}{2m_eR^2}+e\mathcal{E}_{0}Rf(t)\Big[\cos(\beta)\cos(\omega_{L}t)\cos(\varphi
)+\nonumber\\
 &+&\sin(\beta)\sin(\omega_{L}t)\sin(\varphi)\Big],
\end{eqnarray}
where $\ell_z$ is the $z$ component of the orbital angular momentum operator (in units of $\hbar$), whose eigenvectors and eigenvalues are the angular momentum states $\ell_z\mid m\rangle=m\mid\ m\rangle$ with $m=0, \pm 1,\dots$,  and $m_e$ is the electron mass. Solving the time dependent Schr\"odinger equation of the system, we find the expansion coefficient $a_m$ of the wave function $\mid t\rangle$:
\begin{equation} 
\mid t\rangle=\sum_{m=-\infty}^{+\infty} a_m(t)\mid m\rangle
\end{equation}
\noindent where $\mid m\rangle$ are  eigenstates of the free Hamiltonian with energy
\begin{eqnarray}
\label{eq:energy}
\hbar\omega_m=\frac{\hbar^2}{2m_eR^2}m^2.
\end{eqnarray}

From the state$\mid t\rangle$ we can calculate the dipole moment of the system as:
\begin{eqnarray}
\mathcal{\vec{D}}&=&e\vec{r}(t)=\hat{\epsilon}_x\langle t\mid x\mid t\rangle+\hat{\epsilon}_y\langle t\mid y\mid t\rangle=\nonumber\\ &=&\sum_{m=-\infty}^{+\infty}\bigl[\hat{\epsilon}_x\Re\left(a^*_{m-1}a_{m}\right)+\hat{\epsilon}_y\Im\left(a^*_{m}a_{m-1}\right)\bigr]
\end{eqnarray}
where $\vec{r}$ is the position of the electron.
The angular momentum acquired by the electron is:
\begin{equation}
 L_z(t)= \sum_{m=-\infty}^{+\infty}| a_m(t)|^2\hbar m, \label{eq:momentum}
\end{equation}
and the correspondent time averaged angular momentum is:
\begin{eqnarray}
\label{eq:lta}
\langle L_z\rangle=\frac{1}{T}\int_0^TL_z(t)\textbf{d}t.
\end{eqnarray}
From the angular momentum we can define the magnetic momentum as $\vec{m}=\gamma\vec{L}$, with $\gamma$ the gyromagnetic ratio of the electron.
\section{Results}
\begin{figure}
\includegraphics[scale=0.42]{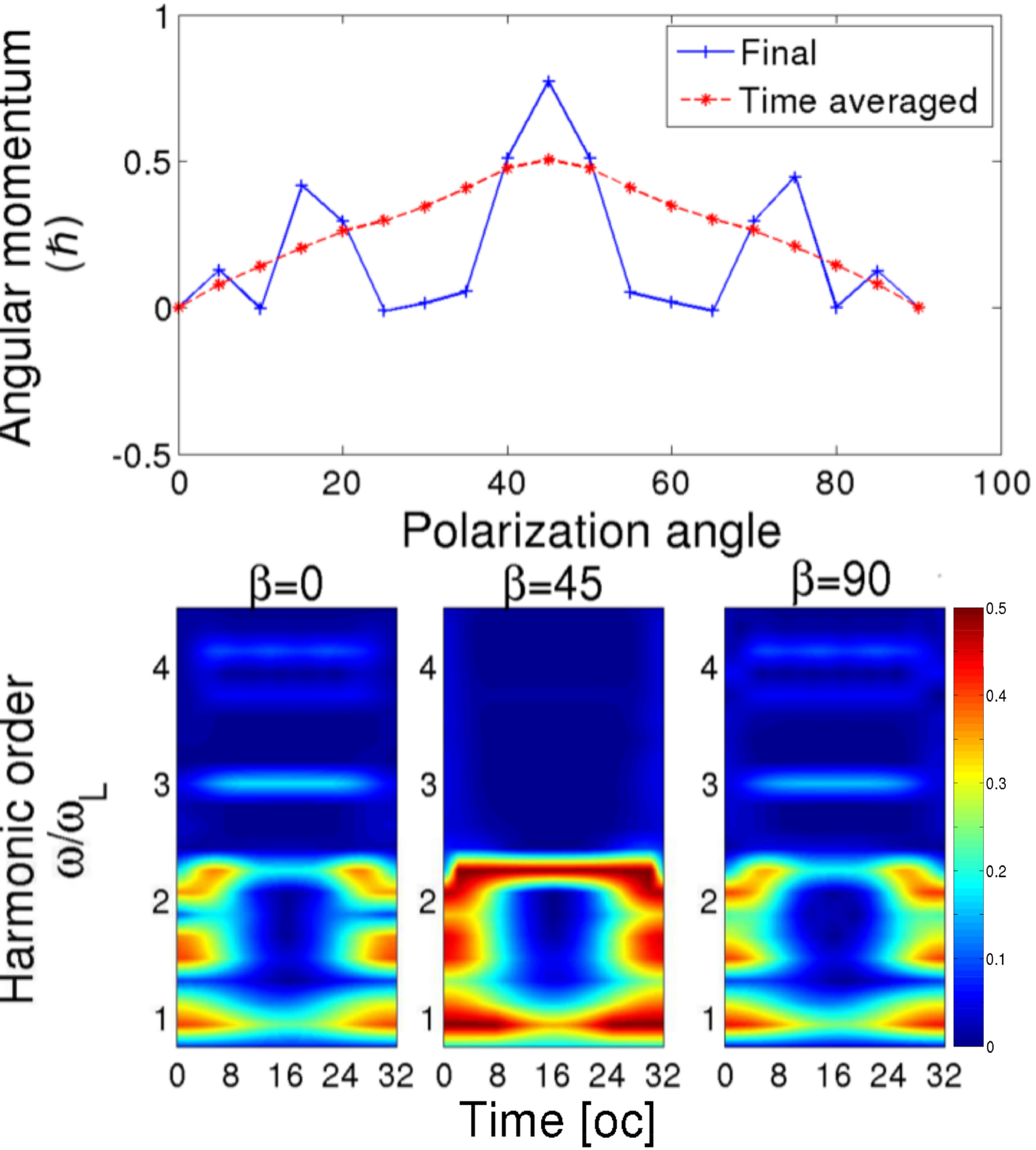}
\caption{Top: Final (solid line) and time averaged (dashed line) angular momentum obtained with $\hbar\omega_L=2$ eV and $I_L=10^{14}$ W/cm$^2$ and R=2.7 $a_0$. Bottom: Morlet wavelet analysis of the dipole moment for $\beta=0^\circ, 45^\circ, 90^\circ$.}
\label{trapzero}
\end{figure}

In our calculations we use different configurations of laser intensity and laser photon energy. In particular we use laser intensities $I_L$ in the range  $10^{10} - 10^{14}$ W$\slash$cm$^2$ and laser energy $\hbar\omega_L$ in the range  $0.1 - 2$ eV, that in wavelength correspond to 12398 and 620 nm,  and a radius between  $R=2.7$ $a_0$, like the radius of the aromatic group, and  $R=100$ $a_0$. In fact large value of radius require small energies of the laser photon and less laser intensity.
Then the dependence upon the radius permits to engineer the nanoring to obtain particular transition energies and to make it flexible for various uses.
In Fig. \ref{trapzero} (top) we show the final  and the time averaged  angular momentum versus the polarization angle  with R=2.7 $a_0$, $\hbar\omega_L=2$ eV and $I_L=10^{14}$ W$\slash$cm$^2$. The time averaged angular momentum is calculated for each polarization angle using the Eq. \ref{eq:lta}. In this simulation we use a laser duration of 32 optical cycles (oc) with a trapezoidal pulse shape. To understand when a signal is present, we performed a Morlet wavelet analysis  on the total spectrum.  
The Morlet mother wavelet is defined as:
\begin{eqnarray}
M(x)=\left(e^{-ix}-e^{-\frac{\sigma_0^2}{2}}\right)e^{-\frac{x^2}{2\sigma_0^2}}
\end{eqnarray}
where $\sigma_0$ is a parameter that indicates the time-frequency resolution of the integration \cite{fiordeluca}. In our calculation, we chose $\sigma_0=6$, that correspond to 6 oscillations of the signal within the Morlet wavelet shape.
We indicate with $H_{1}$ and $H_{II}$, the first two odd harmonics of the spectrum and with $H_{R1}$ and $H_{R2}$ the signals corresponding to the Raman transitions located near the first and the second even harmonics of the HHG spectrum. In fact by the Eq. \ref{eq:energy} we have $\hbar\omega_1=1.9$ eV, $\hbar\omega_2=7.6$ eV, $\hbar\omega_3=17$ eV, then we can define the signal $H_{R1}$ as the transition between the virtual level with energy of $2\hbar\omega_1=3.8$ eV and the ground state, and signal $H_{R2}$ as the combination of two Raman transition: $\hbar\omega_2$ and  $\hbar\omega_2+\hbar\omega_1=9.4$ eV. The Raman transitions depend upon the radius of the nanoring, we chose the value R=2.7 $a_0$ in order make the system comparable to a single cell of graphene or an aromatic group. 

\begin{figure}
\includegraphics[scale=0.5]{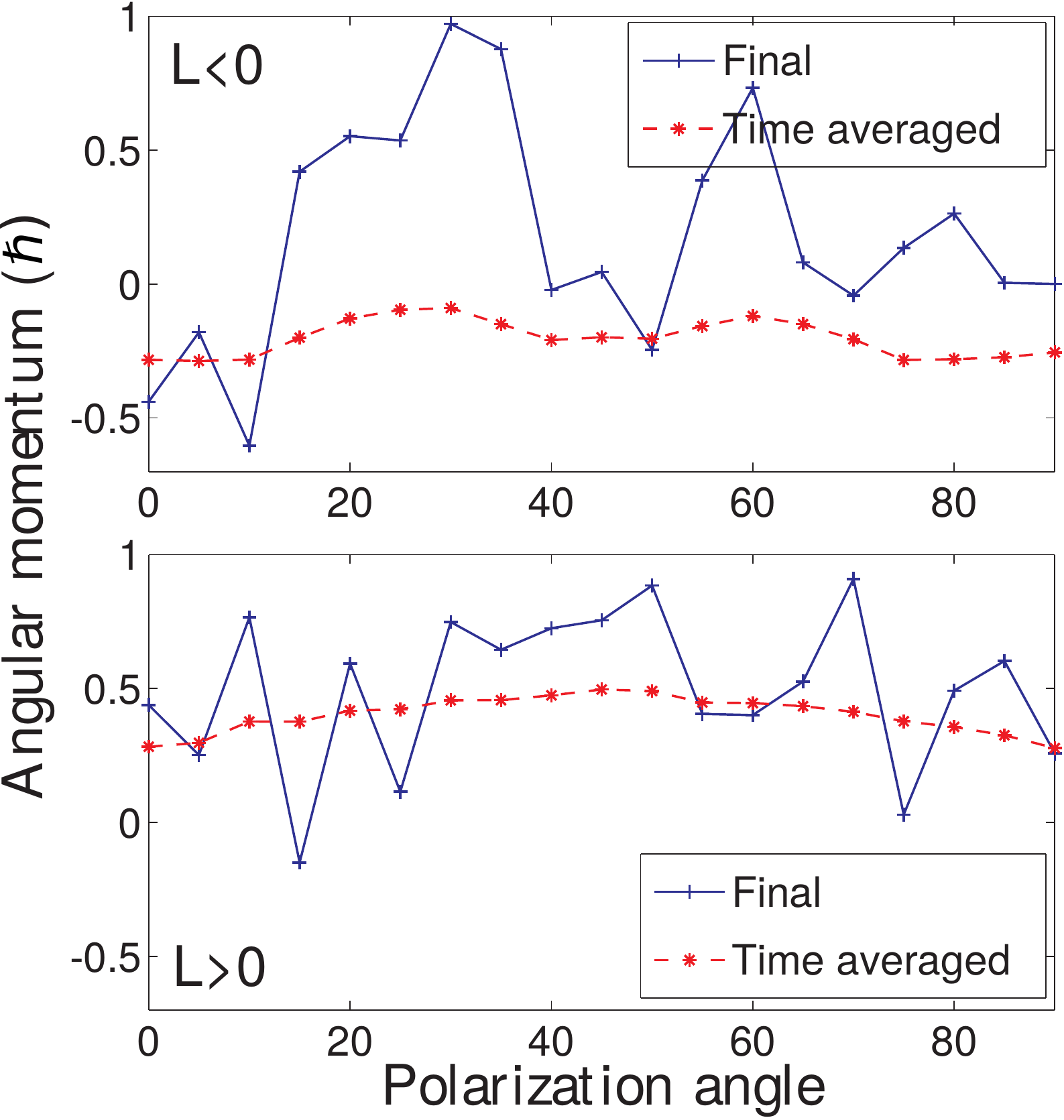}
\caption{Final (solid line) and time averaged (dashed line) angular momentum obtained with $\hbar\omega_L=2$ eV and $I_L=10^{14}$ W/cm$^2$ and R=2.7 $a_0$. Initial angular momentum $L<0$ (top) and $L>0$ (bottom).}
\label{trapL}
\end{figure}

\begin{figure}
\includegraphics[width=\textwidth]{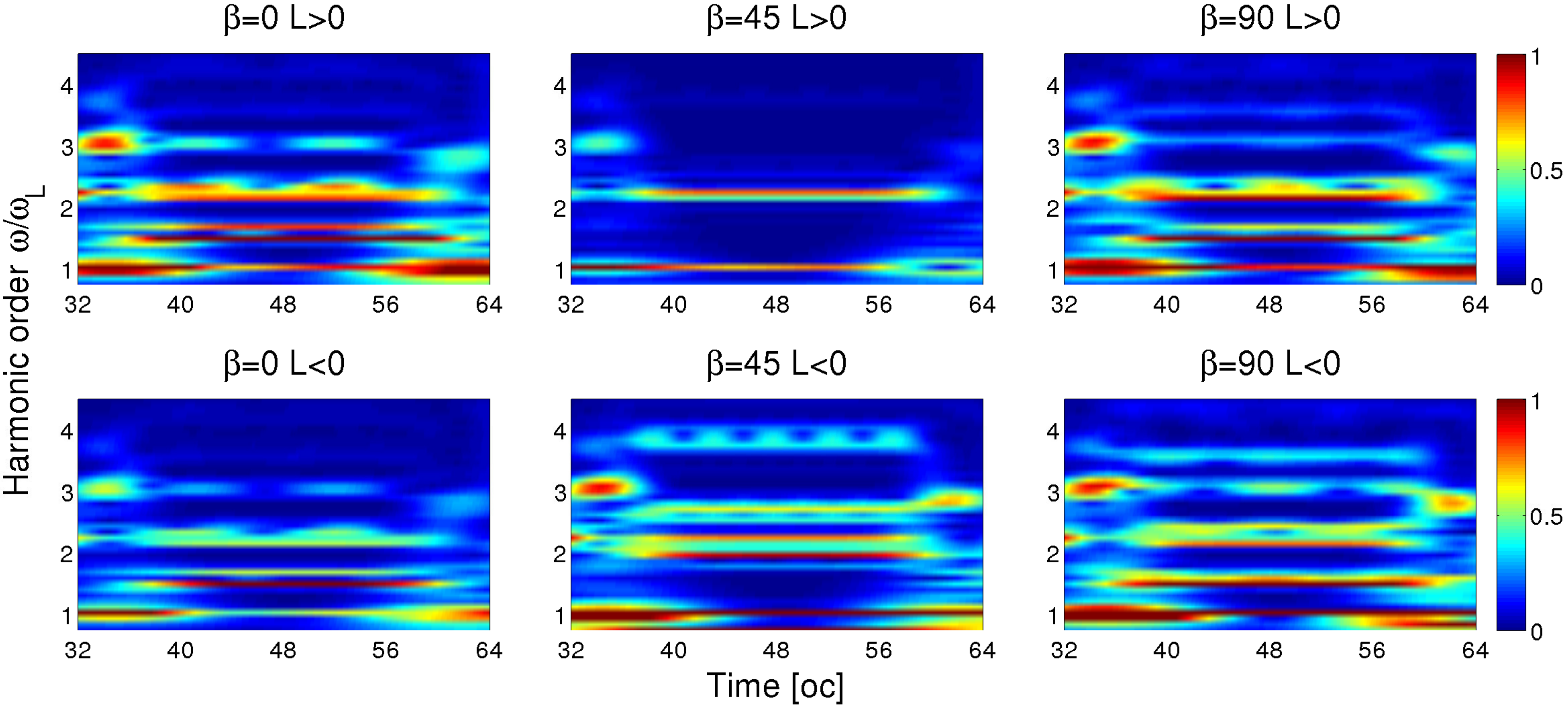}
\caption{Morlet wavelet analysis  of the dipole moment with $\hbar\omega_L=2$ eV and $I_L=10^{14}$ W/cm$^2$ and R=2.7 $a_0$ for $\beta=0^\circ, 45^\circ, 90^\circ$ and an initial angular momentum $L > 0$ (top line) and $L < 0$ (bottom line).}
\label{wavmorl0L}
\end{figure}

 In Fig. \ref{trapzero} (bottom) we show the wavelet analysis for $\beta=0^\circ, 45^\circ, 90^\circ$. In this analysis the line $H_{R1}$ is slightly shifted  up the 4 eV by the presence of a non Raman line. In fact if within the oscillations in  $\sigma_0$ we have several lines, the final value of the wavelet is shifted towards the more intense line. Now we create a truth table where we associate  the values 1 and 0 to the polarization states of the incident laser field and to the presence or absence of signals. In particular we can divide the elliptical polarized laser into two components: $\mathcal{E}_{x}$ parallel to the $x$ axis and $\mathcal{E}_{y}$ parallel to the $y$ axis. When the laser is off, we  have the state $\mathcal{E}_{x}=0$ and $\mathcal{E}_{y}=0$ ($\mathcal{E}_{x,y}=(0,0)$); for $\beta=0^\circ \rightarrow \mathcal{E}_{x,y}=(1,0)$ 
, for $\beta=90^\circ \rightarrow \mathcal{E}_{x,y}=(0,1)$ and for $\beta=45^\circ \rightarrow \mathcal{E}_{x,y}=(1,1)$. We also associate the value 1 when the system presents a final angular momentum $L_z$.
In Tab. \ref{tabzero} we can see that the first odd harmonic and the $H_{R1}$ line behave as a OR logic gate, the second odd harmonic and the $H_{R2}$ line behaves as a XOR logic gate and $L_z$ behave as a AND logic gate.
Now we study the system using two consecutive laser pulses. We use the first laser pulse, circularly polarized, as a pump ($\beta=45^\circ$) and the second laser pulse, elliptically polarized, to probe the system. We make this choice in order to prepare the system with an initial angular momentum. From the dashed line of Fig. \ref{trapzero} we can see that for $\beta=45^\circ$ we have a positive time averaged angular momentum   and the emission of the first  odd harmonic and the $H_{R1}$ line (Fig. \ref{trapzero} (bottom-middle)). We obtain different results varying the sign of the initial angular momentum. In Fig. \ref{trapL} we show the final  and the time averaged angular momentum  with different signs of the initial angular momentum: $L<0$ (top) and $L>0$ (bottom). In this simulation we used a laser duration of 64 oc where in the first 32 oc we prepared the system with an initial angular momentum in order to study the nanoring with different starting conditions. We used a trapezoidal laser shape with 2 oc of ascent and descent laser. In Fig. \ref{wavmorl0L} we show the respective wavelet analysis. If we have a positive  initial angular momentum we obtain a BUFFER for the first odd harmonic, the $H_{R1}$ line, and for the final angular momentum, a XOR logic gate for the second odd harmonic and a RESET for the $H_{R2}$ line; if the initial angular momentum is negative, we obtain a BUFFER  for the first odd harmonic and the $H_{R1}$ line,  an AND logic gate for the $H_{R2}$ line and an OR for the second odd harmonic. These results are listed in the truth table  of Tab. \ref{tabL}. 

\begin{figure}
\includegraphics[width=\textwidth]{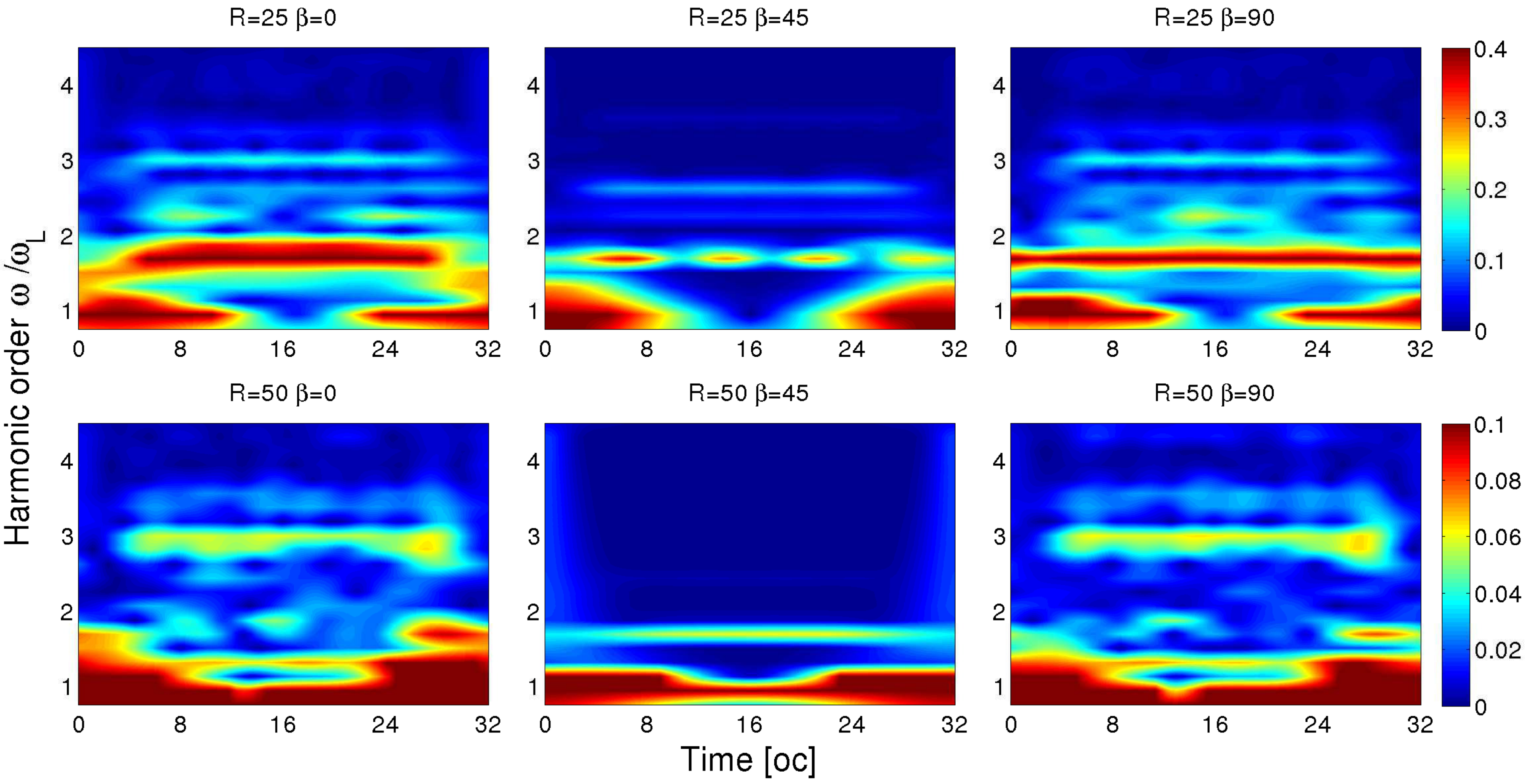}
\caption{Morlet wavelet analysis  of the dipole moment for $\beta=0^\circ, 45^\circ, 90^\circ$, a radius of $R=25$ (top line) and $R=50$ $a_0$ (bottom line), a energy of laser photon of 0.1 eV and an intensity of  $10^{10}$ W$\slash$cm$^2$.}
\label{Rwav}
\end{figure}

We also decrease the laser intensity and the energy of the photon lasera and increase the radius up to 100 $a_0$. In fact with large radii we obtain the same results but using a laser intensity of 10$^{10}$ W$\slash$cm$^2$ and a laser photon of 0.1 eV.
 These new parameters are preferable for the constructions of logic port because there is not the risk of destroying the object.  In Fig. \ref{Rwav} we show the wavelet transform using a radius of $R=25$ and $R=50$ $a_0$ respectively, a laser intensity of $10^{10}$ W$\slash$cm$^2$ and a laser photon of 0.1 eV and without an initial angular momentum. We can see that the structure of the emitted lines is the same but  less defined. This because the harmonic spectrum present several non Raman lines that shift and enlarge the lines.

Now we show how nanorings can be arranged to form logic circuits. Examples of logical circuits are the half and the full adder.

\begin{figure}
\includegraphics[scale=1.6]{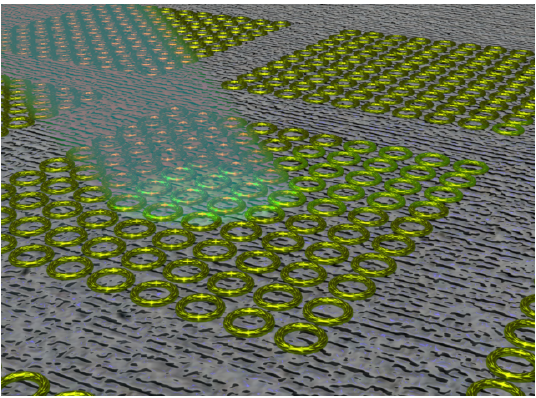}
\caption{Schematic representation of  array of nanorings used to store informations. When one array is driven by the laser field, it creates a magnetic moment that represents a bit. }
\label{ringarray}
\end{figure}

The half adder is a digital  electronic component that has in input two bits ($\mathcal{E}_{x}$ and $\mathcal{E}_{y}$ in our case) and give in output their sum ($S$) and their  carry ($C$). We can use the nanoring without initial angular momentum as an half adder using as output the second odd harmonic,  the $H_{R2}$ line and $L_z$ (Tab. \ref{tab:halfadder}). Combining two nanorings, we have the possibility to make a full adder, where we  use the carry of a previous summation. In fact we can use the harmonics obtained by one nanoring as input for a second nanoring et cetera. In this way we obtain a logic circuit with a set of concatenated nanorings.
We now suggested a way of using the nanoring to store information. When a nanoring is driven by a circular polarized laser field, the electron will round on it with a circular motion. This movement will generate a current $I$. If we call $\vec{S}$ the surface of the nanoring, the magnetic moment generated by the motion of the electron will be:
\begin{equation}
 \vec{m}=I\oint\textbf{d}\vec{S},
 \end{equation} 
 or, in terms of the angular momentum: $\vec{m}=\gamma\vec{L}$.
 If we have a planar array of nanorings and if $\vec{S}$ is the surface of the laser spot containing several nanorings, the magnetic momentum generated will be proportionally to the summation of the the contributions of each nanoring within the laser spot surface (Fig. \ref{ringarray}).  Then we can use the nanoring to store information with the magnetic momentum.
 
When the nanoring  is driven by a laser field in the state (1,1), it acquire an angular momentum and we obtain $L_z=1$. Then if we use another laser pulses with inverse circular polarization, we obtain an angular momentum state $L_z=0$; we can consider the angular momentum like a pseudo-spin.  

These simple behaviours make the nanoring an interesting object to store information.

\section{Conclusions}
In this article we can consider the nanorings a real alternative to modern logic components thanks to their size and speed. 
We investigated the possibility to use nanorings driven by a laser field to make logic circuits. In particular we used the emitted signals and the final angular momentum of the nanoring to create logic gate that can be used to make logic operations. In fact we noticed the possibility to construct the XOR, OR and AND logic gates and use them to make a half and full adder. Combining  two or more nanorings, we can obtain a full adder, but the presence of XOR and AND logic gates give us the possibility to make a Toffoli gate. The Toffoli gate is an universal reversible logic gate that permits to construct  any reversible circuit. It has 3 bit  inputs ($a, b$ and $c$) and outputs and realizes the function $c$ XOR $(a$ AND $b)$. In fact  if we set the first two bits, the Toffoli gate inverts the third bit, otherwise all bits stay the same. Then it is possible to use the nanorings as a Toffoli gate and create any logic circuit.

 We also can use the angular momentum acquired by the electron in the nanoring to store information. In fact we can consider the final angular momentum like a pseudo-spin  that can be reversed by changing the direction of circular polarization of the incident laser field. The entire process, including  harmonics generation, Raman transitions, and angular momentum,  takes about $10^{-15}$ seconds.
In addition we can create cells of nanoring arrays and store informations on it using the laser to generate a magnetic moment. In this paper we showed several simulations  to study the system with different parameters, such as the radius, the energy of the laser photon and the intensity of the laser. In fact if we enlarge the radius and decrease the energy laser photon, we obtain the same results. If we think to construct a nanoring with a radius of 25 or 50 $a_0$, we can use a laser photon with energy $\sim 0.1$ eV and a laser intensity of $\sim 10^{10}$ W$\slash$cm$^2$.

\section{Acknowledgment}
We  thank Franco Persico for his reading of the manuscript, comments and suggestions.
\bibliography{bibring}{}
\bibliographystyle{unsrt}

\begin{table}
 \begin{tabular}{|c|c|c|c|c|c|}
\hline 
$\mathcal{E}_{x,y}$ & $H_I$ & $H_{II}$ & $H_{R1}$ & $H_{R2}$ & $L_z$ \\ 
\hline 
0 0&  0 & 0 & 0 & 0 &0\\ 
\hline 
1 0&  1 & 1 & 1 & 1 &0\\ 
\hline 
0 1&  1 & 1 & 1 & 1 &0\\ 
\hline 
1 1&  1 & 0 & 1 & 0 &1\\ 
\hline 
 $L=0$&OR & XOR & OR& XOR& AND \\ 
\hline 
\end{tabular}
\caption{\label{tabzero}Truth table: in input we have the laser states $\mathcal{E}_{x,y}$ and in output we have the first two odd harmonics, the Raman transitions and the final angular momentum.}
\end{table}

\begin{table}
\begin{tabular}{|c|c|c|c|c|c|c|}
 \hline 
 $\mathcal{E}_{x,y}$ &  $H_{1}$&$H_{II}$& $H_{R1}$ & $H_{R2}$ & $L_z$ \\ 
 \hline 
 0 0&  1 & 0 & 1 & 0 & 1 \\ 
 \hline 
 1 0&  1 & 1 & 1 & 0 & 1 \\ 
 \hline 
 0 1&   1 & 1 & 1 & 0 & 1 \\ 
 \hline 
 1 1&  1 & 0& 1 & 0 & 1 \\ 
 \hline 
 L$>$0 &BUFFER &XOR &BUFFER&RESET& BUFFER \\ 
 \hline
  \multicolumn{6}{l}{ }\\
 \hline 
 $\mathcal{E}_{x,y}$  &$H_{1}$&$H_{II}$& $H_{R1}$ & $H_{R2}$ & $L_z$ \\ 
 \hline 
 0 0&   1 & 0 &1 & 0 & 1 \\ 
 \hline 
 1 0&   1 & 1 & 1 & 0 & 1 \\ 
 \hline 
 0 1&  1 & 1 & 1 & 0 & 0 \\ 
 \hline 
 1 1&  1 & 1 & 1& 1 & 0 \\ 
  \hline 
 L$<$0 &BUFFER&OR&BUFFER&AND& // \\ 
 \hline 
 \end{tabular} 
\caption{\label{tabL}Truth table: in input we have the laser states $\mathcal{E}_{x,y}$ with a initial positive angular momentum (top) and with a negative initial angular momentum (bottom). In output we have the first two odd harmonics, the Raman transitions  and the final angular momentum.}
\end{table}

\begin{table}
\begin{tabular}{|c|c|c|}
\hline 
$\mathcal{E}_{x,y}$ & Sum & Carry \\ 
\hline 
0 0 & 0 & 0 \\ 
\hline 
1 0 & 1 & 0 \\ 
\hline 
0 1 & 1 & 0\\ 
\hline 
1 1 &0 & 1 \\ 
\hline 
\end{tabular} 
\caption{\label{tab:halfadder}Truth table of the half adder: in input we have the laser states $\mathcal{E}_{x,y}$ and in output we have the values of  the second odd harmonic and the $H_{R2}$ line for the sum, and the values of $L_z$ for the carry.}
\end{table}
\end{document}